\RequirePackage{ifpdf}
\ifpdf 
\documentclass[pdftex]{sigma}
\else
\documentclass{sigma}
\fi

\numberwithin{equation}{section}

\begin{document}


\renewcommand{\PaperNumber}{013}

\FirstPageHeading

\renewcommand{\thefootnote}{$\star$}

\ShortArticleName{Towards Unifying Structures in Higher Spin Gauge Symmetry}

\ArticleName{Towards Unifying Structures \\
in Higher Spin Gauge Symmetry\footnote{This
paper is a contribution to the Proceedings of the Seventh
International Conference ``Symmetry in Nonlinear Mathematical
Physics'' (June 24--30, 2007, Kyiv, Ukraine). The full collection
is available at
\href{http://www.emis.de/journals/SIGMA/symmetry2007.html}{http://www.emis.de/journals/SIGMA/symmetry2007.html}}}

\Author{Anders K.H. BENGTSSON}
\AuthorNameForHeading{A.K.H. Bengtsson}

\Address{School of Engineering, University College of Bor{\aa}s, All\' egatan 1, SE-50190 Bor\aa s, Sweden}
\Email{\href{mailto:anders.bengtsson@hb.se}{anders.bengtsson@hb.se}}
\URLaddress{\url{http://www.adm.hb.se/~abe/}}

\ArticleDates{Received November 01, 2007, in f\/inal form January
25, 2008; Published online February 04, 2008}

\Abstract{This article is expository in nature, outlining some of the many still incompletely understood features of higher spin f\/ield theory. We are mainly considering higher spin gauge f\/ields in their own right as free-standing theoretical constructs and not circumstances where they occur as part of another system. Considering the problem of introducing interactions among higher spin gauge f\/ields, there has historically been two broad avenues of approach. One approach entails gauging a non-Abelian global symmetry algebra, in the process making it local. The other approach entails deforming an already local but Abelian gauge algebra, in the process making it non-Abelian. In cases where both avenues have been explored, such as for spin 1 and 2 gauge f\/ields, the results agree (barring conceptual and technical issues) with Yang--Mills theory and Einstein gravity. In the case of an inf\/inite tower of higher spin gauge f\/ields, the f\/irst approach has been thoroughly developed and explored by M.~Vasiliev, whereas the second approach, after having lain dormant for a long time, has received new attention by several authors lately. In the present paper we brief\/ly review some aspects of the history of higher spin gauge f\/ields as a backdrop to an attempt at comparing the gauging vs. deforming approaches. A common unifying structure of strongly homotopy Lie algebras underlying both approaches will be discussed. The modern deformation approach, using BRST-BV methods, will be described as far as it is developed at the present time. The f\/irst steps of a formulation in the categorical language of operads will be outlined. A~few aspects of the subject that seems not to have been thoroughly investigated are pointed out.}


\Classification{70S10; 70S20; 70G60; 70H45; 81T13; 81T70}

\vspace{-2mm}

\section{Introduction}\label{sec:Introduction}
Back in 1978, Fang and Fronsdal, in a paper discussing the deformation theoretic approach to deriving a self-interacting massless spin-2 f\/ield theory, proposed what they called a ``generali\-zed Gupta program'' to the ef\/fect of deriving self-interactions for higher spin gauge f\/ields \cite{FangFronsdal1979}. Fronsdal apparently soon realized that such a program was not likely to succeed for any single spin $s$. In a conference proceedings from 1979 he suggested that, what we now refer to as an {\it infinite tower} of higher spin gauge f\/ields, is needed. Indeed he wrote \cite{Fronsdal1979conf}:
``{\it For spins exceeding 2 it would seem to be very difficult to find a non-Abelian gauge algebra without including all spins or at least all integer spins. Thus the question calls for a single, unifying gauge algebra for all integer spins.}''

That something like this could be the case can surmised from an unsophisticated generalization from spin 1 and 2. For spin 1, commuting two ``gauge generators'' $\xi_a T^a$ and $\eta_a T^a$ yields a new transformation of the same form. Likewise, by analogy, for spin 2, commuting $\xi^\mu\partial_\mu$ and $\eta^\mu\partial_\mu$, yields an object of the same form. The obvious generalization to spin 3 would be commuting objects of the form $\xi^{\mu\nu}_aT^a\partial_\mu\partial_\nu$. However, this operation does not close \cite{AKHB1985}. In the language of deformation theory, we discover an obstruction. The form of the non-closure terms in the commutator suggests the introduction of spin 5 gauge transformations~\cite{Burgers1985thesis}. The need for a  ``single unifying gauge algebra for all integer spins'' can therefore be discerned.

Furthermore, the step to thinking in terms of inf\/inite component objects $\Phi$ in some way maintaining f\/ields of all spin, is now short. In the paper referred to above, Fronsdal outlined one such approach which will be described below.

Thus, adding in the fact that the massless representations of the Poincar{\'e} group can be reali\-zed on symmetric tensor f\/ields (possibly modulo certain traces), these considerations together strongly hint at some kind of expansion
\begin{gather*}
\Phi=\phi B+\phi^\mu B_\mu+\phi^{\mu\nu} B_{\mu\nu}+\phi^{\mu\nu\rho} B_{\mu\nu\rho}+\cdots,
\end{gather*}
where the set $\{B_{\mu_1\mu_2\ldots\mu_s}\}_{s=0}^\infty=\{B_{(s)}\}_{s=0}^\infty$ can be thought of as an inf\/inite basis or an inf\/inite set of generators. Many such proposals can be found in the literature, indeed, this simple idea seems to be reinvented ever and anon.
The question arises as to whether the $B$'s obey an algebra, for instance
\begin{gather*}
[B_{(s)},B_{(t)}]=\sum_{i,j}^{s,t}c^{ij}B_{(i,j)}
\end{gather*}
in a vague, but hopefully suggestive notation, or if they simply commute. Of course, higher order brackets might occur, or be necessary, at least the Jacobiator must be considered. This opens up the whole f\/ield of abstract algebra.

Clearly, without guiding principles, we're groping about in the dark.  However, after thirty years of groping about by many authors, the territory is glowing feebly. The purpose of the present paper is certainly not to attempt a full review, rather we will try to further illuminate some parts of the terrain. Lacking guiding principles, our work will very much be playing around with equations.

Since in my opinion, not very much is gained by working in higher dimensions, I will work in four spacetime dimensions unless otherwise stated.

\section{Higher spin bases and generators}\label{sec:HSBG}
In this section, without any claims of being exhaustive, we will look at a few examples of higher spin bases/generators. Perhaps the simplest one, and certainly one that obviously might come to mind upon trying to generalize spin 1 and 2, is a derivative basis already alluded to in the introduction.

\subsection{Back to zero}
When two inf\/initesimal transformations $\delta_\xi$ and $\delta_\eta$ in some space are performed consecutively, the result is in general dif\/ferent depending on the order in which the transformations are done. This is one reason why it is interesting to study the commutator $[\delta_\xi,\delta_\eta]$ of the transformations. This could be anything, but it seems that the interesting cases arise when this dif\/ference can itself be expressed as a new transformation of the same type as the ones that are commuted. In any case, once the commutator is singled out as an interesting object of study, the Jacobi identities are forced upon the theory. As they are simply syntactical consequences of writing out $[[\delta_{\xi_1},\delta_{\xi_2}],\delta_{\xi_3}]+\mbox{cyc.\,perm.}$ and assuming that the inf\/initesimal operations $\delta_\xi$ are associative, there is no way to get around them. Thus the semantics of the theory must obey them, i.e.\ whatever results from explicitly calculating the commutator brackets, it must obey the Jacobi identities.

In general, brackets $[\cdot,\cdot]$ are used as a notation for operations that are not commutators def\/ined in terms of an underlying associative multiplication. In such cases the Jacobiator $[[\cdot,\cdot],\cdot]+\mbox{cyc.\,perm.}$ might only be zero up to higher order brackets, or {\it homotopies}. In any case, independent axioms are needed for the Jacobiator.

\subsection{Derivative basis}
Even on the weak inductive base step of comparing the sequence of natural numbers $s=1,2,3,\ldots$ to the known sequence of spin 1 and 2 gauge parameters $\xi^a$, $\xi^\mu$, it could have been guessed that the higher spin continuation of this sequence should be something like $\xi^{a\mu_1\mu_2}$, $\xi^{\mu_1\mu_2\mu_3},\ldots$ with some internal index $a\in\{1,\ldots,n\}$ for odd spin and $s-1$ spacetime indices $\mu_i$ for spin $s$. The construction in 1983 of the cubic covariant vertex in Minkowski spacetime~\cite{BerendsBurgersvanDam1984} for spin~3 and the light-front arbitrary $s$ cubic vertices \cite{BBB1983a} would have corroborated such a guess.

Now let us see what, if anything, can be surmised from taking a derivative basis seriously. To begin with, assume that a spin 3 gauge generator is actually given by an object of the form $\xi^{\mu\nu}_aT^a\partial_\mu\partial_\nu$. Commuting two of these yield
\begin{gather}
[\xi^{\mu\nu}_{1a}T^a\partial_\mu\partial_\nu,\xi^{\rho\sigma}_{2b}T^b\partial_\rho\partial_\sigma]=
\xi^{\mu\nu}_{1a}\xi^{\rho\sigma}_{2b}[T^a,T^b]\partial_\mu\partial_\nu\partial_\rho\partial_\sigma+
2(\xi^{\mu\nu}_{1a}\partial_\mu\xi^{\rho\sigma}_{2b}-\xi^{\mu\nu}_{2a}\partial_\mu\xi^{\rho\sigma}_{1b})
T^aT^b\partial_\nu\partial_\rho\partial_\sigma\nonumber\\
\phantom{[\xi^{\mu\nu}_{1a}T^a\partial_\mu\partial_\nu,\xi^{\rho\sigma}_{2b}T^b\partial_\rho\partial_\sigma]=}{} +(\xi^{\mu\nu}_{1a}\partial_\mu\partial_\nu\xi^{\rho\sigma}_{2b}
-\xi^{\mu\nu}_{2a}\partial_\mu\partial_\nu\xi^{\rho\sigma}_{1b})T^aT^b\partial_\rho\partial_\sigma.\label{eq:CommutingSpin3Generators}
\end{gather}

The f\/irst term is easily interpreted as a spin 5 transformation by taking $[T_a,T_b]=f_{ab}^{\,\,\,\,c}T_c$
as we would be prejudiced to require of the matrices $T$. However, then the last term becomes problematic since it clearly must be a new spin 3 transformation. It the seems (provisionally) that we need a weaker set of equations for products of matrices, namely
\begin{gather}\label{eq:MatrixEquations}
T_aT_b=g_{ab}^{\,\,\,\,c}T_c,
\end{gather}
with no conditions on the coef\/f\/icients $g_{ab}^{\,\,\,\,c}$.

Next, the second term of (\ref{eq:CommutingSpin3Generators}) which corresponds to a spin 4 transformation, implies that also the even higher spin ($s>2$) f\/ields must carry internal indices. However, the light-front cubic interaction terms for even higher spin require no anti-symmetric internal indexing, indeed in that case the interactions would vanish. On the other hand, commuting two spin 4 generators of the form $\xi^{\mu\nu\rho}$, yield spin 3 generators without internal matrices $T$.

These considerations show that in order to make a simpleminded scheme like the one described here work, every higher spin f\/ield must be expanded over a set of matrices $T$ with a product satisfying equation (\ref{eq:MatrixEquations}) as well as having a component along a unit matrix $E$, or
\begin{gather*}
\phi=\varphi^aT_a+\varphi E.
\end{gather*}

Taking $E$ to be $T_0$ we let the gauge f\/ields be valued in a Clif\/ford algebra with basis elements~$T_a$, $a\in\{0,\ldots,N\}$. The equation (\ref{eq:MatrixEquations}) can then be satisf\/ied with suitable choices for the $g_{ab}^{\,\,\,\,c}$ (such bases can be built using {\it matrix units} \cite{FuchsScweigertBook}).

Tentatively, the would-be gauge algebra is generated by operators
\begin{gather*}
\Xi=\xi^a T_a+\sum_{k=1}^\infty\xi^{a\mu_{i_1}\ldots\mu_{i_k}}T_a\partial_{\mu_{i_1}}\cdots\partial_{\mu_{i_k}},
\end{gather*}
acting as
\begin{gather*}
\delta_\xi\Phi=[\Xi,\Phi].
\end{gather*}

The crucial question is of course whether operators of this form generate a structure that can be considered to be a gauge algebra? This was studied in \cite{BB1986}. There it was shown that the operators $\Xi$ obey the Jacobi identity (which they do since they associate as a consequence of the associativity of the matrices $T$ and the Leibniz rule for the derivatives $\partial$) and that structure functions can be extracted according to the following scheme.

Using the Jacobi identity
\begin{gather*}
[\delta_\lambda,\delta_\xi]\Phi=[\Lambda,[\Xi,\Phi]]-[\Xi,[\Lambda,\Phi]]=[[\Lambda,\Xi],\Phi],
\end{gather*}
we f\/ind that the commutator of two transformations can be written
\begin{gather*}
[\delta_\lambda,\delta_\xi]\Phi=\delta_{[\lambda,\xi]}\Phi=\delta_\omega\Phi,
\end{gather*}
with structure thus extracted as $\Omega=[\Lambda,\Xi]$.

There are however some problems with this approach, the most serious that it does not seem to reproduce the lowest order spin 3 structure equations $[\lambda,\xi]^a_{\mu\nu}$ of \cite{BerendsBurgersvanDam1984,Burgers1985thesis} (BBvD). I write {\it seem}, because as far as I know, this has not really been investigated. Reconsidering the whole setup, it is clear that, had the derivative basis scheme worked, we would have had a f\/ield independent higher spin gauge algebra. According to the analysis of (BBvD), they f\/ind f\/ield dependence once one goes beyond spin 2. They then perform a general analysis of the gauge algebra problem and f\/ind that the spin 3 gauge algebra cannot close on spin 3 f\/ields. The non-closure terms however have a form that suggests that they correspond to spin 5 gauge transformations. Unfortunately it is not known whether in this way introducing a tower of higher spin gauge f\/ields, the full algebra would become f\/ield independent. Clearly, there must be more interesting structure to extract here. There is a recent paper that alludes to this \cite{Bekaert2007}.

\subsection{Fronsdal's symplectic basis}
In the conference proceedings referred to in the introduction \cite{Fronsdal1979conf}, Fronsdal brief\/ly outlines an attempt to set up a non-Abelian higher spin gauge algebra. He considers a cotangent phase space over spacetime with the usual symplectic structure and coordinates $(x^\mu,\pi_\nu)$. Then he considers a set of traceless symmetric gauge parameters $\{\xi^{\mu_1\dots\mu_{s-1}}\}_{s=0}^\infty$ and the corresponding set of gauge f\/ields $\{\phi^{\mu_1\dots\mu_{s}}\}_{s=1}^\infty$. These are collected into formal power series
\begin{gather*}
\Xi(\pi,x)=\sum_{s=1}^\infty(\pi^2)^{1-s/2}\pi_{\mu_1}\cdots\pi_{\mu_{s-1}}\xi^{\mu_1\dots\mu_{s-1}},\nonumber\\
\Phi(\pi,x)=\pi^2+\sum_{s=1}^\infty(\pi^2)^{1-s/2}\pi_{\mu_1}\cdots\pi_{\mu_{s}}\phi^{\mu_1\dots\mu_{s}}.
\end{gather*}

Transformations can now be calculated using the Poisson bracket $\{x_\mu,\pi_\nu\}=\eta_{\mu\nu}$ between $x_\mu$ and $\pi_\nu$
\begin{gather*}
\delta_\Xi\Phi=\{\Xi,\Phi\}.
\end{gather*}

The presence of the $\pi^2$ term in $H$ is needed in order to get the free theory transformations
\begin{gather*}
\{\Xi,\pi^2\}=2\sum_{s=1}^\infty(\pi^2)^{1-s/2}\pi_{\mu_1}\cdots\pi_{\mu_{s}}\partial^{\mu_1}\xi^{\mu_2\dots\mu_{s}}.
\end{gather*}
Again, the Jacobi identity allows us to extract structure equations
\begin{gather*}
[\delta_\Lambda,\delta_\Xi]\Phi=\{\Lambda,\{\Xi,\Phi\}\}-\{\Xi,\{\Lambda,\Phi\}\}=\delta_{\{\Lambda,\Xi\}}\Phi.
\end{gather*}
And once again we would get a f\/ield independent gauge algebra. Note that this construction is a form of Schouten brackets.

\subsection{Oscillator basis}\label{subsec:OscBas}
Formally introducing a covariant harmonic oscillator pair $(\alpha_\mu,\alpha_\mu^\dagger)$ with the usual commutator $[\alpha_\mu,\alpha_\mu^\dagger]=\eta_{\mu\nu}$ and a vacuum $|0\rangle$ satisfying $\alpha_\mu|0\rangle=0$, one could consider collecting all higher spin gauge f\/ields in the expansion
\begin{gather}\label{eq:OscillatorExpansionFields}
|\Phi\rangle=(\phi +\phi^\mu\alpha_\mu^\dagger+\phi^{\mu\nu}\alpha_\mu^\dagger\alpha_\nu^\dagger+
\phi^{\mu\nu\rho}\alpha_\mu^\dagger\alpha_\nu^\dagger\alpha_\rho^\dagger+\cdots)|0\rangle.
\end{gather}

This yields a concrete calculational scheme and one could hope to be able to rig some kind of kinetic operator $K$ which when acting on the f\/ield $|\Phi\rangle$, generate f\/ield equations for the component f\/ields. This can indeed be done as is well known. In practice it is done using BRST methods. Several equivalent such formulations were discovered in the mid 1980's \cite{SiegelZwiebach1987,OuvryStern1987a,AKHB1987a,Meurice1988}.

Formulations of this form has subsequently been rediscovered, developed and extended by a~few research groups during the last ten years \cite{Labastida1989,PashnevTsulaia1997,PashnevTsulaia1998a,FranciaSagnotti2002a}. Of course, expansions like (\ref{eq:OscillatorExpansionFields}) f\/irst occurred in the bosonic string theory, and the BRST formulations was partly inspired by string f\/ield theory and by taking the limit $\alpha '\rightarrow\infty$ (or the zero tension limit) \cite{AKHB1987a}. The zero tension limit has been much studied lately (see for example \cite{Sundborg2001a,Bonelli2003a,SagnottiTsulaia2004a}). For some recent reviews of these developments as well as more complete sets of references see for example \cite{FranciaSagnotti2006rw,Bianchi2004a} and also \cite{SagnottiSezginSundell2005,FranciaMouradSagnotti2007a} which contain many references as well interesting discussions of these topics. This is closely tangled up with ideas about holography and the AdS/CFT duality \cite{Maldacena1997a} in a~way that seems not fully understood. It is thus obvious that higher spin gauge f\/ields occur in very many contexts. They are a prevalent feature of any model of extended and/or composite relativistic systems in various background geometries and dimensions. For discussions of these topics, see~\cite{SezginSundell2002a,BeisertBianchiMoralesSamtleben2004a}. But rather than study them in these contexts, I will continue to discuss higher spin symmetry and higher spin gauge theory as a free-standing theoretical construct. Thus let us return to the oscillator-BRST formulation and brief\/ly review the simplest model.

Consider therefore a phase space spanned by bosonic variables $(x_\mu,p_\mu)$ and $(\alpha_\mu,\alpha^{\dagger}_\mu)$ and ghost variables $(c^+,b_+)$, $(c^-,b_-)$ and $(c^0,b_0)$ with commutation relations
\begin{gather*}
[x_{\mu},p_{\nu}]=i\eta_{\mu\nu},\qquad[\alpha_{\mu},\alpha_{\nu}^{\dagger}]=\eta_{\mu\nu},
\qquad\{c^+,b_+\}=\{c^-,b_-\}=\{c^0,b_0\}=1.
\end{gather*}

The ghosts have the following properties under Hermitian conjugation
\begin{gather*}
(c^-)^\dagger=c^+,\qquad(b_-)^\dagger=b_+,\qquad(c^0)^\dagger=c^0,\qquad(b_0)^\dagger=b_0.
\end{gather*}

The doubly degenerate vacuum states $|+\rangle$, $|-\rangle$ are annihilated by the operators $\alpha_\mu$, $c^-$, $b^+$ while the degeneracy is given by
\begin{gather*}
b_0|+\rangle=0,\qquad b_0|-\rangle=|+\rangle,\qquad c^0|-\rangle=0,\qquad c^0|+\rangle=|-\rangle.
\end{gather*}

These equations relating the vacua, then implies that either one of the two vacua must be odd. I will choose $|-\rangle$ Grassmann odd.

The higher spin f\/ields are collected into the ket $|\Phi\rangle$ with expansion
\begin{gather*}
|\Phi\rangle=\Phi(p)|+\rangle+F(p)c^+b_{-}|+\rangle+H(p)b_-)|-\rangle,
\end{gather*}
where $\Phi(p)$ contains the symmetric higher spin gauge f\/ields, and $F(p)$ and $H(p)$ are certain auxiliary f\/ields. These f\/ields are further expanded in terms of the oscillators
\begin{gather*}
\Phi=\Phi_0+i\Phi^\mu\alpha^\dagger_\mu+\Phi^{\mu\nu}\alpha^\dagger_\mu\alpha^\dagger_\nu+\cdots,\\
F=F_0+iF^\mu\alpha^\dagger_\mu+F^{\mu\nu}\alpha^\dagger_\mu\alpha^\dagger_\nu+\cdots, \\
H=H_0+iH^\mu\alpha^\dagger_\mu+H^{\mu\nu}\alpha^\dagger_\mu\alpha^\dagger_\nu+\cdots.
\end{gather*}

The gauge parameters are collected in
\begin{gather*}
|\Xi\rangle=(\xi_0-i\xi^\mu\alpha^\dagger_\mu+\xi^{\mu\nu}\alpha^\dagger_\mu\alpha^\dagger_\nu+\cdots)b_-|+\rangle.
\end{gather*}

The BRST operator $Q$ is expressed in terms of the generators
\begin{gather*}
G_0=\tfrac 12 p^2,\qquad G_-=\alpha\cdot p,\qquad G_+=\alpha^{\dagger}\cdot p,
\end{gather*}
spanning the simple algebra
\begin{gather*}
[G_-,G_+]=2G_0,
\end{gather*}
with all other commutators zero.

In terms of these generators, the BRST operator reads
\begin{gather*}
Q=c^0G_0-c^+G_+ -c^-G_- -2c^+c^-b_0.
\end{gather*}

The action
\begin{gather}\label{eq:FreeAction}
A=\langle\Phi|Q|\Phi\rangle,
\end{gather}
is invariant under the gauge transformations,
\begin{gather}\label{eq:FreeGaugeTransformation}
\delta_\Xi|\Phi\rangle=Q|\Xi\rangle,
\end{gather}
as is the f\/ield equation
\begin{gather}\label{eq:FreeFieldEquation}
Q|\Phi\rangle=0.
\end{gather}

When expanding the equations (\ref{eq:FreeAction}), (\ref{eq:FreeGaugeTransformation}) and (\ref{eq:FreeFieldEquation}), everything works out nicely for the component f\/ields, except for the fact that the theory contains auxiliary f\/ields which cannot be solved for without introducing a further constraint. This constraint is applied to both the f\/ield and the gauge parameter
\begin{gather*}
T|\Phi\rangle=0,\qquad T|\Xi\rangle=0,
\end{gather*}
where $T$ is the operator
\begin{gather*}
T=\tfrac 12 \alpha\cdot\alpha+b_+c^-.
\end{gather*}

When expanded, the constraint equations yield the double tracelessness constraint for component f\/ields of spin $s\geq 4$ and the tracelessness constraint for the corresponding component gauge parameters. The constraints are needed in order to get the correct number of physical degrees of freedom. Including the constraints, this formulation precisely reproduces the Fronsdal equations \cite{Fronsdal1978} for higher spin gauge f\/ields.

The free f\/ield theory is however still gauge invariant without imposing these constraints, and as discussed in \cite{AKHB2007a}, the constraints can actually be discarded with. The presence of the extra f\/ields are then instead controlled by two global symmetries of the theory \cite{AKHB1988}. Other approaches to circumventing the tracelessness constraints have been proposed \cite{FranciaSagnotti2003a}. With some extra work this theory can be formulated in terms the Batalin--Vilkovisky f\/ield/antif\/ield formalism. We will return to this issue in Section~\ref{sec:RotBRST-BVMA}.

Perhaps just as interesting as the technical reconstruction of Fronsdal's free higher spin theory in terms of the BRST formalism, is the hint at an underlying mechanical model. In the f\/irst two schemes outlined above, it is not clear what the bases {\it are}. In the f\/irst case, generalizing from spin 1 $\xi_aT^a$ and spin 2 $\xi^\mu\partial_\mu$, (where for spin 2 we really should consider a full Poincar{\'e} generator $\xi^\mu\partial_\mu+{1\over 2}\xi^{ij}M_{ij}$) it is clear that we have no conceptual understanding what the higher derivative operators actually do. Perhaps somewhat vaguely we could say that we are re-using spacetime tangent space, not introducing anything new. The same goes for Fronsdal's approach. Although clever, it is clear that the scheme again is just re-using spacetime, but now its cotangent superstructure.

In contrast to this, introducing oscillators do add a something new, indeed new dimensions apart from spacetime itself. It is possible construct mechanical models that produces the f\/irst class constraints behind the above BRST construction of the f\/ield theory~\cite{AKHB1987b, HenneauxTeitelboim1989a}. During the 1960's and 70's there were other, parallel approaches to strong interactions apart from string theory (before QCD and asymptotic freedom swept the table). These also involved inf\/inite towers of massless particles/f\/ields and ideas about underlying composite mechanical models \cite{EighthNobelSymposium,CasalbuoniLonghi1975}, though at that time the interest was not primarily in the higher spin gauge f\/ields.

\section{Deformation vs. gauging}\label{sec:DvsG}
There are two main approaches to introducing interactions into a massless free spin 1 theory. Either gauge a global but non-Abelian Lie algebra, or deform local but Abelian gauge symmetries. It is of some interest to reconsider these approaches in order to get a hold on what is involved in generalizing to all spins.

One problem with the approaches described so far, apart from the fact that very few solid positive results on interactions have been achieved, is that they suf\/fer from a weak conceptual underpinning. In my opinion this is also true for the much more successful Vasiliev approach, although in this case there are quite a few technical circumstances that lend some basic strength to the approach. We will now try to understand this in a simpleminded straightforward way, and in the process clarify some of the connections between the AdS and the Minkowski space (MiS) formulations of higher spin theory.

The Vasiliev approach is well described in the literature so I will not attempt yet another review, but instead focus some relevant points, perhaps in a somewhat idiosyncratic way.

\subsection{What are the higher spin algebras?}
The Vasiliev approach to higher spins seems simple enough. Just take a higher spin algebra and gauge it. Now, from where do we get the higher spin algebras? That's also easy, just take the algebra $so(3,2)$ or any of its higher dimensional and/or supersymmetric versions and form powers of its generators to any order. What results is an inf\/inite dimensional associative algebra. The rest is technicalities. These are amply covered in the literature, what we want to do here is something more basic. The question is: why would this work? And why would it work in AdS and not in MiS.

Let $T^a$ be the generators of an $N$-dimensional Lie algebra. Forming free powers of these generators and collecting these into the set
\begin{gather}\label{eq:InfiniteDimAssociativeAlgebra}
{\bf A}=\{T^{a_1}T^{a_2}\cdots T^{a_{i-1}}T^{a_i}\}_{i=1}^\infty
\end{gather}
we can promote {\bf A} to an associative algebra through the universal enveloping algebra construction. This involves forming equivalence classes of generators by modding out by the elements of the ideal generated by elements of the form $[T^a,T^b]$ and then taking care of automorphisms.

Non-compact Lie algebras have oscillator representations and these are easily seen to be inf\/inite dimensional. Thus when we use a non-compact Lie algebra such as $so(3,2)$ to promote it to an inf\/inite dimensional associative algebra using the construction (\ref{eq:InfiniteDimAssociativeAlgebra}), what we are really doing is to use its inf\/inite-dimensional representations as a kind of basis for an inf\/inite dimensional algebra. It is worthwhile to do this in some detail for the basic case of $so(3,2)$.

\subsubsection*{The higher spin algebra $\boldsymbol{hso(3,2)}$}

Let us then consider a pair of independent oscillators $(a,a^\dagger)$ and $(b,b^\dagger)$ with $[a,a^{\dagger}]=1$ and $[b,b ^{\dagger}]=1$. Using these we can build low dimensional Lie algebras. Thus for the compact algebra $su(2)$ we def\/ine the generators
\begin{gather*}
J_+= b^{\dagger}a,\qquad
J_-=a^{\dagger}b,\qquad
J_3=\tfrac 12 (b^{\dagger}b-a^{\dagger}a).
\end{gather*}
spanning
\begin{gather*}
[J_3,J_+]=J_+,\qquad [J_3,J_-]=-J_-,\qquad [J_+,J_-]=2J_3.
\end{gather*}

To the $su(2)$ generators we can add the number operator
\begin{gather*}
E=\tfrac 12 (b^{\dagger}b+a^{\dagger}a+1),
\end{gather*}
which commutes with the rest of the generators. This turns $su(2)$ into $u(2)$.

Next, using just one oscillator, we can build the non-compact $sp(2)$ algebra
\begin{gather*}
S_+ =\tfrac 12 a^{\dagger}a^{\dagger},\qquad
S_- = \tfrac 12 aa,\qquad
S_3 = \tfrac 12 \big(a^{\dagger}a+\tfrac 12\big).
\end{gather*}

The algebra is
\begin{gather*}
[S_3,S_+]=S_+,\qquad [S_3,S_-]=-S_-,\qquad [S_+,S_-]=-2S_3.
\end{gather*}

Starting with $sp(2)$ it is clear that we get an inf\/inite dimensional representation over an oscillator ground state $|0\rangle$ with $a|0\rangle=0$. Also, forming powers of the generators $S_+$, $S_-$, $S_3$ we get an inf\/inite dimensional algebra represented over this same representation space. This is all well known including the necessary brush up details.

Now what happens in the $su(2)$ case? Def\/ine the double ground state $|0_a,0_b\rangle\doteq|0_a\rangle|0_b\rangle$ with the obvious properties. For short we just write $|0,0\rangle$. Excited states are written $|n_a,n_b\rangle$.

The Fock space ${\bf F}_{su(2)}$ built upon the spin 0 ground state $|0,0\rangle$ gets a natural grading by spin. Using standard notation $|jm\rangle$ for angular momentum states we get
\begin{gather}
{\bf F}_{su(2)} = \{|0,0\rangle\}\cup\{(|0,1\rangle,|1,0\rangle\}\cup\{|0,2\rangle,|1,1\rangle,|2,0\rangle)\}\cup\cdots\nonumber\\
\phantom{{\bf F}_{su(2)}}{} =\bigcup_{n=0,1,2,\ldots}\big(\cup_{n_a+n_b=n}\{|n_a,n_b\rangle\}\big)
=\bigcup_{j=0,{1\over2},1,\ldots}(\oplus_{m=-j}^{m=j}|jm\rangle).\label{eq:SU2FockSpace}
\end{gather}

We can now consider powers of the $su(2)$ generators, say $(J_+)^p$ and $(J_-)^q$. Then acting with these operators on spin ${n\over 2}$ states $|n_a,n_b\rangle$ we either get zero or a new spin ${n\over 2}$ state (when $p,q\leq n$), i.e.\ we stay within the same subspace, or we get zero (when $p,q>n$). Furthermore, since all states are eigenstates of $J_3$, powers of $J_3$ also stay within the same spin ${n\over 2}$ subspace.  So in this way we cannot build an inf\/inite dimensional algebra based on the compact $su(2)$ algebra that acts transitively on the full Fock space of states.

Let us now turn to the non-compact algebra $so(3,2)$. Its ten generators splits into three components according to $so(3,2)\mapsto g^{-1}\oplus g^{0}\oplus g^{+1}$ where
\begin{gather*}
g^{-1} = \{L_-^-,L_+^-,L_3^-\},\qquad
g^0= \{E,J_+,J_-,J_3\},\qquad
g^{+1}= \{L_+^+,L_-^+,L_3^+\}.
\end{gather*}

The $g^0$ generators are precisely the already def\/ined $u(2)$ generators. The rest are expressed in terms of the oscillators as
\begin{gather*}
L_-^-=aa,\qquad L_+^-=bb,\qquad L_3^-=ab,\qquad
L_+^+=a^\dagger a^\dagger,\qquad L_-^+=b^\dagger b^\dagger,\qquad L_3^+=a^\dagger b^\dagger.
\end{gather*}

Clearly, the $g^{-1}$ are lowering operators and the $g^{+1}$ are raising operators, and the overall structure of the algebra is
\begin{gather*}
[g^m,g^m]\subseteq g^{m+n},\qquad m,n=\{\pm1,0\}.
\end{gather*}

Using the raising and lowering operators we can now step up and down in the full Fock space~${\bf F}_{su(2)}$. This Fock space is actually the ${\rm Di}\oplus{\rm Rac}$ Fock space of states $|e,j\rangle$ with ground state $|{1\over 2},0\rangle$ and with the dispersion equation $e=j+{1\over 2}$  \cite{Dirac1963,FlatoFronsdal1978a,FlatoFronsdal1980a,GunaydinSacliogu1982a,GunaydinPreprint1989}. The subset of states $\{|n_a,n_b\rangle:n_a+n_b=n\}$ has $e={1\over 2}(n_a+n_b+1)$ and $m={1\over 2}(n_a-n_b)$, where as in (\ref{eq:SU2FockSpace}), $m$ is the~$J_3$ quantum number ranging between $-j$ and $j$.

Finally, modulo technical details, the higher spin algebra $hso(3,2)$ is built by simply taking all positive powers of the $so(3,2)$ generators. It becomes an inf\/inite-dimensional Lie algebra transitively represented on the ${\rm Di}\oplus{\rm Rac}$ weight space (alternatively the Fock space ${\bf F}_{su(2)}$).

So what are the higher spin algebras? It seems that in four spacetime dimensions, the simplest higher spin algebra is that of an inf\/inite-dimensional transformation algebra acting on the direct sum of all representations of the 3D angular momentum algebra. The crucial point is that the algebra connects states of dif\/ferent spin. What we get is a concrete realization of the universal enveloping algebra of $so(3,2)$.

A question is now, can this be done for the Poincar{\'e} algebra? Analyzing how the Poincar{\'e} algebra is embedded in $so(3,2)$ it becomes clear that the $so(3,2)$ raising and lowering operators~${L^{\pm}_a}$  are linear combinations of Poincar{\'e} translations and boosts with the deformation parameter $\epsilon={1\over\sqrt\Lambda}$ as coef\/f\/icient ($a$ is a space index in the set $\{1,2,3\}$ or $\{+,-,3\}$)
\begin{gather*}
L^{\pm}_a=\epsilon P_a\pm i L_{a0}.
\end{gather*}

Thus when the Wigner--In\"on\"u contraction $\Lambda\rightarrow 0$ is performed, the raising and lowering operators break up, and we lose much of the interesting AdS structure. Still there are remnants of the structure in MiS as is apparent from in the BRST oscillator expansion (\ref{eq:OscillatorExpansionFields}) of higher spin gauge f\/ields. An interesting line of research would be to extend the MiS Lorentz algebra $so(3,1)$ to an inf\/inite algebra.

\subsection{Gauging}
Whereas the gauge theory approach to spin 1 interactions is fairly straightforward, indeed it is the paradigmatic example, gauging approaches to interacting spin 2 has always been plagued by conceptual (and technical) dif\/f\/iculties. As a backdrop, let us brief\/ly run through the spin 1 case.

We have a non-Abelian Lie algebra represented by matrices $T_a$
\begin{gather*}
[T_a,T_b]=f_{ab}^{\,\,\,\,c}T_c,
\end{gather*}
and a vector of matter f\/ields $\varphi$ transforming in some representation
\begin{gather*}
\delta\varphi=\epsilon\varphi=\epsilon^aT_a\varphi,
\end{gather*}
and a matter Lagrangian which is invariant under these transformations. Making the para\-me\-ters $\epsilon$ local: $\epsilon(x)$, we f\/ind the problem that spacetime derivatives of the f\/ields transform in-homogeneously
\begin{gather*}
\delta(\partial_\mu\varphi)=\partial_\mu(\delta\varphi)=\epsilon^aT_a\partial_\mu\varphi+(\partial_\mu\epsilon^a)T_a\varphi.
\end{gather*}

The remedy is introducing new gauge f\/ields $A_\mu=A_\mu^aT_a$ transforming as
\begin{gather*}
\delta A_\mu=[\epsilon,A_\mu]-\partial_\mu\epsilon.
\end{gather*}

Then the covariant derivative $D_\mu=\partial_\mu+A_\mu$ transforms homogeneously. Replacing ordinary derivatives with covariant ones then restores invariance to the matter Lagrangian. However, in our present context, the matter Lagrangian is just a crutch, what we are after is non-linear dynamics for the gauge f\/ield itself. The solution is quite simple and beautiful. Commuting covariant derivatives we f\/ind the f\/ield strengths $F_{\mu\nu}$ according to
\begin{gather}\label{eq:FieldStrengths}
F_{\mu\nu}=[D_\mu,D_\nu]=\partial_\mu A_\nu-\partial_\nu A_\mu+[A_\mu,A_\nu].
\end{gather}

These transform as
\begin{gather*}
\delta F_{\mu\nu}=[\epsilon,F_{\mu\nu}],
\end{gather*}
and it f\/inally transpires that a non-linear, gauge-invariant Lagrangian can be written
\begin{gather*}
{\cal L}=-{1\over 4g^2} F_{\mu\nu}^aF^{\mu\nu}_a,
\end{gather*}
from which we read of\/f that the interacting theory is a deformation of free spin 1 gauge theory running up to cubic and quartic order in the interaction. (Note that in the above brief outline the gauge coupling constant $g$ has been absorbed into the f\/ields and parameters, it can be restored by the substitutions $A\mapsto gA,\epsilon\mapsto g\epsilon$.)

What is the strong point of this? Clearly the fact that the correct non-linearity is forced upon us through the equation (\ref{eq:FieldStrengths}). It is like following a recipe. The weak point is f\/inding the correct spin 1 Lagrangian, it looks simple enough here, but this step does not generalize easily to higher spin.

In the case of spin 2, the trouble starts at the very f\/irst step, choosing the gauge group. Having General Relativity in the back of our minds, we think about the local tangent spaces. In a frame, or vierbein, formulation this becomes very concrete as local Poincar{\'e} transformations. But we are running somewhat ahead of ourselves. Start with an (active) inf\/initesimal Poincar{\'e} transformation
\begin{gather*}
\delta x^\mu=\epsilon^\mu_{\,\,\,\nu} x^\nu+\epsilon^\mu,
\end{gather*}
and correspondingly, a set of matter f\/ields $\varphi$ transforming as
\begin{gather}\label{eq:PoincareOnMatterField}
\delta\varphi=\tfrac 12 \epsilon^{ij}S_{ij}\varphi-\epsilon^\mu\partial_\mu\varphi.
\end{gather}

Many things have slipped in here. In the f\/irst term we have changed indices from (curved) spacetime indices $\mu$, $\nu$ to tangent space Lorentz indices $i$, $j$. So far we can see that we are really just generalizing the spin 1 recipe in a straightforward manner. The role of the $T_a$ matrices are now played by the Lorentz $so(3,1)$ matrices $S_{ij}$. In the second term we see that corresponding to the translation part of the Poincar{\'e} transformations, we get generators that are spacetime derivatives. Anyone who has played around with these equations for a while is likely to suf\/fer from some confusion, and it gets worse. As soon as the parameters $\epsilon^{ij}$ and $\epsilon^\mu$ are made into local functions of the coordinates $x^\mu$ the distinction between local translations and local Lorentz transformations becomes blurred. Translations with a local $\epsilon^\mu(x)$ already contains all local coordinate transformations generated by the vector f\/ield $\epsilon^\mu(x)\partial_\mu$. However, if the local Lorentz transformations $\epsilon^{ij}S_{ij}$ are discarded, the transformations on the f\/ields (\ref{eq:PoincareOnMatterField}) become ambiguous.

Correspondingly, there are various approaches in the literature, gauging the Lorentz \linebreak group~\cite{Kibble}, gauging the translations only, or gauging the full Poincar{\'e} group. A few more papers from the heydays of this approach to gravity are \cite{Utiyama,MacDowellMansouri1977,MansouriChang,Grensing,HehlHeydeKerlick}.

Following the review article \cite{KibbleStelle1986} and proceeding to gauge the full Poincar{\'e} group, we again f\/ind that the derivate of a matter f\/ield transforms inhomogeneously
\begin{gather*}
\delta\partial_\mu\varphi=\tfrac 12 \epsilon^{ij}S_{ij}\partial_\mu\varphi-\epsilon^\nu\partial_\nu\partial_\mu\varphi+
\tfrac 12 (\partial_\mu\epsilon^{ij})S_{ij}\varphi-(\partial_\mu\epsilon^\nu)\partial_\nu\varphi.
\end{gather*}

The third term can be taken care of by introducing a gauge f\/ield $\omega^{ij}_\mu$ with an inhomogeneous transformation term $-\partial_\mu\epsilon^{ij}$, and correspondingly we have a covariant derivative
\begin{gather*}
\nabla_\mu=\partial_\mu+\tfrac 12 \omega^{\,\,\,ij}_\mu.
\end{gather*}
The last term must be treated in a dif\/ferent way since it involves derivatives $\partial_\mu$ instead of matrices. Here we do a multiplicative gauging introducing the vierbeins $e^{\,\,\mu}_i$
\begin{gather*}
D_i=e^{\,\,\mu}_i\nabla_\mu.
\end{gather*}

The rest of the story from here on involves commuting the covariant derivatives to f\/ind curvature $R_{ij}^{\;\;\;kl}$ and torsion tensors $T_{ij}^{\;\;k}$. Due to the multiplicative gauging, the curvature tensor is second order in spacetime derivatives (in contrast to the spin 1 f\/ield strengths which are f\/irst order in derivatives), and this is ref\/lected in the gravity Lagrangian being expressed as
\begin{gather*}
{\cal L}\simeq R_{ij}^{\;\;\;ij}.
\end{gather*}

The theory so obtained is almost Einstein gravity, the dif\/ference is the presence of tor\-sion~\cite{Kibble,HehlHeydeKerlick}.

This very (indeed) brief review of the gauging approaches to spin 1 and 2, shows disturbing dif\/ferences between the two cases. There certainly isn't any standard recipe to generalize. In a very loose language we could say that Yang--Mills theory is basically very algebraic whereas Gravity is very geometric. One way to iron out the dif\/ferences would be to geometrize Yang--Mills, this is precisely what one does in the modern f\/iber bundle approach. Another way would be to make Gravity more algebraic.

As we saw, one source to the problems came from the translation part of the Poincar{\'e} group which gives rise to derivatives as gauge generators. However in the AdS group $SO(3,2)$ the translation generators $P_\mu$ are just a subset of the $so(3,2)$ generators according to $P_\mu=\sqrt \Lambda M_{\mu 4}$ with $[P_\mu,P_\nu]=-i\Lambda M_{\mu\nu}$. So one could attempt to set up an $SO(3,2)$ gauge theory of gravity. This was done by Kibble and Stelle \cite{StelleWest1980,KibbleStelle1986} (see also \cite{MacDowellMansouri1977}). Standard Einstein gravity is recovered upon spontaneously breaking $SO(3,2)$ to Poincar{\'e} corresponding to a Wigner--In\"on\"u contraction $\Lambda\rightarrow0$. This could be thought of as making gravity more algebraic. The Vasiliev approach to higher spin follow this road.

\subsection{The core of the Vasiliev theory}
What the Vasiliev theory of higher spins does is basically that it extends the gauge algebra $so(3,2)$ to the higher spin algebra $hso(3,2)$ (and correspondingly in other dimensions with or without supersymmetry) and then by generalizing the techniques of  MacDowell and Mansouri~\cite{MacDowellMansouri1977}) and Stelle and West \cite{StelleWest1980}, it manages to derive f\/ield equations for interacting higher spin gauge f\/ields coupled to gravity and Yang--Mills. This is a very impressive achievement and it involves a lot of technical details. Still it has so far not been possible to write down a unifying Lagrangian for this theory, and partly for this reason it must be said that the approach is still not entirely understood.

The f\/ield equations of the Vasiliev approach are written as a generalization of the Maurer--Cartan equations for a Lie algebra (for good reviews, see \cite{BekaertCnockaertIazeollaVasiliev2005,JohanEngqvist} which we follow here)
\begin{gather}\label{eq:MaurerCartan1}
R^a\doteq d\sigma^a+\tfrac 12 f_{bc}^{\,\,\,\,a}\sigma^b\wedge\sigma^c=0,
\end{gather}
and where the $\sigma^a$ are 1-forms on the Lie group manifold.

The Jacobi identity follows from the integrability conditions $dR^a=0$ using $dd=0$. This formulation is completely equivalent to the more common formulation in terms of vector f\/ields (or generators) $T_a$ satisfying the usual Lie brackets $[T_a,T_b]=f_{ab}^{\;\;\;c}T_c$ and with the inner product $\langle\sigma^a,T_b\rangle=\delta^a_ b$.

The Maurer--Cartan equations are invariant under gauge transformations
\begin{gather*}
\delta\sigma^a=D\epsilon^a\doteq d\epsilon^a+f_{bc}^{\,\,\,\,a}\sigma^b\wedge\epsilon^c.
\end{gather*}

This is then generalized by introducing, a possibly inf\/inite, collection of $(p+1)$-form curvatu\-res~$R^\alpha$ def\/ined as
\begin{gather}\label{eq:MaurerCartan2}
R^\alpha\doteq d\Sigma^a+F^\alpha(\Sigma^\beta)=0,
\end{gather}
where
\begin{gather}\label{eq:DefineStructureMC}
F^\alpha(\Sigma^\beta)\doteq\sum_{k=1}^\infty f_{\;\;\beta_1\dots\beta_k}^{\alpha}\Sigma^{\beta_1}\wedge\cdots\wedge\Sigma^{\beta_k},
\end{gather}
where the $\Sigma^\beta$ are $p$-forms on the Lie group manifold. Here $f_{\beta_1\dots\beta_k}^{\alpha}$ are the structure constants of a, possibly inf\/inite-dimensional, algebra. The form degrees must match, i.e.\ $\sum\limits_{i=1}^k p_{\beta_i}=p_\alpha+1$.

Now the integrability conditions read $dF^\alpha(\Sigma^\beta)=0$, or
\begin{gather}\label{eq:IntegrabilityConditionMC}
F^\beta\wedge{\partial F^\alpha\over\partial\Sigma^\beta}=0,
\end{gather}
yielding generalized Jacobi identities for the structure constants $f_{\beta_1\dots\beta_k}^{\alpha}$. The gauge transformations also generalize to
\begin{gather*}
\delta\Sigma^\alpha=d\epsilon^\alpha-\epsilon^\beta\wedge{\partial F^\alpha\over\partial\Sigma^\beta},
\end{gather*}
under which the generalized curvatures of (\ref{eq:MaurerCartan2}) stays invariant.

This kind of structure was initially developed in the context supergravity and termed Cartan Integrable Systems \cite{DAuriaFre1982a}, later renamed as Free Dif\/ferential Algebra in accordance with the corresponding mathematical constructs \cite{Sullivan1977}.

The actual f\/ield equations are formulated in terms of two dif\/ferential forms, one zero-form $\Phi=\Sigma^0$ and one one-form $A=\Sigma^1$ which in their turn are inf\/inite expansions in terms of oscillators, much like the expansions above, but with important dif\/ferences. Vasiliev's expansions are in terms of both creation and annihilation operators and will therefore span inf\/inite-dimensional algebras (given some restrictions), indeed the higher spin algebras.

This is the abstract scheme employed in the Vasiliev approach. Of course, we haven't even touched on the mass of technicalities involved in actually setting up higher spin gauge theory in this framework. Some helpful papers are \cite{BekaertCnockaertIazeollaVasiliev2005,JohanEngqvist,Vasiliev2005a} where also many further references can be found. Here we will restrict ourselves to pointing out how strongly homotopy Lie algebras emerge.

Consider the operator
\begin{gather}\label{eq:KOperator}
K=F^\alpha(\Sigma){\partial\over \partial\Sigma^\alpha}.
\end{gather}
Then calculating the square of this operator we get
\begin{gather*}
K^2=\tfrac 12 K\wedge K=\left(F^\beta\wedge{\partial F^\alpha\over\partial\Sigma^\beta}\right){\partial\over\partial\Sigma^a}=0,
\end{gather*}
which is zero by (\ref{eq:IntegrabilityConditionMC}). From here the step to the algebraic structure of a $L_\infty$ algebra is short. Already in the generalization of~(\ref{eq:MaurerCartan1}) to~(\ref{eq:MaurerCartan2}) and the def\/inition (\ref{eq:DefineStructureMC}) we see a hint of the higher order brackets of an sh-Lie algebra. This becomes more clear if we write the Maurer--Cartan equations in an abstract, coordinate-free way as $d\sigma=-{1\over 2}[\sigma,\sigma]$. Generalizing this recklessly, we would consider multibrackets $[\Sigma,\Sigma,\ldots,\Sigma]$ and the Jacobi identities for the ordinary Lie bracket~$[\cdot,\cdot]$ would generalize to the sh-Lie identities. Let us step back and do this in some more detail.

\subsection{Strongly homotopy Lie algebras and higher order brackets}
There are a few variants of the basic def\/initions of strongly homotopy Lie algebras in the literature (see for example \cite{LadaStasheff1993a,LadaMarkl1995a}), but the following, mildly technical, is suf\/f\/icient for our purpose to bring out the connection to both the Vasiliev formalism and to the product identities of the BRST approach described below.

\subsubsection*{Def\/inition}
Consider a ${\bf Z}_2$ graded vector space $V=V_0\oplus V_1$ over some number f\/ield, and denote the elements by $x$. The grading is given by $\varrho$ with $\varrho(x)=0$ if $x\in V_0$ and $\varrho(x)=1$ if $x\in V_1$. $V$ is supposed to carry a sequence of $n$-linear products denoted by brackets. The graded  $n$-linearity is expressed by
\begin{gather*}
[x_1,\ldots,x_n,x_{n+1},\dots,x_m]=(-)^{\varrho(x_n)\varrho(x_{n+1})}[x_1,\ldots,x_{n+1},x_n,\dots,x_m],\\
\lbrack x_1,\ldots,a_n x_n+b_n x'_n\,\ldots,x_m\rbrack\nonumber\\
\qquad{}= a_n(-)^{\iota(a_n,n)}\lbrack x_1,\ldots,x_n,\ldots,x_m\rbrack+b_n(-)^{\iota(b_n,n)}\lbrack x_1,\ldots,x'_n,\ldots,x_m\rbrack, 
\end{gather*}
where $\iota(a_n,n)={\varrho(a_n)(\varrho(x_1)+\cdots +\varrho(x_{n-1})}$.

The def\/ining identities (or ``main'' identities) for the algebra are, for all $n\in{\bf N}$
\begin{gather}\label{eq:SHDefiningEquations}
\sum_{{k=0}\atop{l=0}}^{k+l=n}\sum_{\pi(k,l)}\epsilon(\pi(k,l))\lbrack\lbrack x_{\pi(1)},\ldots,x_{\pi(k)}],x_{\pi(k+1)},\ldots,x_{\pi(k+l)}\rbrack=0,
\end{gather}
where $\pi(k,l)$ stands for $(k,l)$-shuf\/f\/les. A $(k,l)$-shuf\/f\/le is a permutation $\pi$ of the indices $1,2,\dots$, $k+l$ such that $\pi(1)<\cdots<\pi(k)$ and $\pi(k+1)<\cdots<\pi(k+l)$. $\epsilon(\pi(k,l))$ is the sign picked up during the shuf\/f\/le as the points $x_i$ with indices $0\leq i\leq k$ are taken through the points $x_j$ with indices $k+1\leq j\leq l$. This is just the normal procedure in graded algebras.

The bracket (or braces) notation is common. Note that the $n$-ary brackets are all independent and abstract, not to be thought of as deriving from some underlying product (just as the abstract Lie bracket $[\cdot,\cdot]$ need not derive from a product). As discussed in~\cite{Voronov2003a}, other types of gradings can be considered. The ${\bf Z}_2$ grading employed here can be thought as providing room for a BRST-type algebra with even f\/ields and odd gauge parameters. In the BRST-BV reformulation~\cite{AKHB2007a} (see Section~\ref{sec:RotBRST-BVMA} of the present paper) the mechanical and f\/ield theory ghosts conspire to make all objects even (f\/ields, anti-f\/ields, ghosts and anti-ghosts) so in that case we can in fact do with an ungraded algebra. In order to accommodate the FDA of Vasiliev, a grading with respect to form degree should be def\/ined.

Running the risk of pointing out something that is obvious, let us note that even in the case where we have an underlying associative product, such as when we work with matrix algebras, nothing prevents us from calculating higher order brackets and see what we get. What we would get is roughly linear combinations of elements in the vector space that the algebra is built upon, and these linear combinations would def\/ine for us generalized structure coef\/f\/icients. Such higher order Lie algebras has been investigated in \cite{AzcrragaBueno1997a}. They obey generalized Jacobi identities similar in structure to the sh-Lie identities (of which they are a special case), but then as a consequence of the underlying associative product.

Next following~\cite{Voronov2003a} we can introduce generating functions for the sh-Lie structure. As pointed out there, due to the symmetry and linearity properties, the full sh-Lie structure is f\/ixed by the algebra of even and coinciding elements, $\Sigma$ say. We can indeed think of the $\Sigma$ as vectors in one huge graded vector space, for example graded by form degree, corresponding to the forms of the Vasiliev theory. Next we write equation (\ref{eq:DefineStructureMC}) in terms of brackets
\begin{gather*}
K=\sum_{k\geq 0}{1\over k!}\underbrace{[\Sigma,\Sigma,\ldots,\Sigma]}_{k}.
\end{gather*}
$K$ is itself an element of the vector space and we can consider it as a formal vector f\/ield
\begin{gather*}
K=F^\alpha(\Sigma){\partial\over\partial\Sigma^\alpha},
\end{gather*}
which is what we had before (\ref{eq:KOperator}). But now computing the square $K^2$ using its def\/inition in terms of the brackets, we get Jacobiators
\begin{gather*}
J^n(\Sigma,\ldots,\Sigma)=\sum_{l=0}^{n}{n!\over l!(n-l)!}[\,\underbrace{[\Sigma,\ldots,\Sigma]}_{n-l},\underbrace{\Sigma,\ldots,\Sigma}_{l}\,],
\end{gather*}
for even elements $\Sigma$. In terms of these Jacobiators the ``generalized Jacobi identities'' or simply ``main'' identities from the def\/inition of an sh-Lie algebra become $J^n=0$ for all $n$, or
\begin{gather*}
J=\sum_{n\geq 0}{1\over n!}J^n=K^2=0.
\end{gather*}

This, admittedly hand-waving, argument shows that there is a connection between sh-Lie algebras and FDA's. This is often alluded to in the literature (see for example \cite{BekaertCnockaertIazeollaVasiliev2005} but I haven't seen any rigorous proof. But there cannot really be any doubt that the technical details can be supplied to make the connection precise. According to \cite{Vasiliev2005a}, any FDA can be obtained from an sh-Lie algebra by f\/ixing the form degree $p_\alpha$ of the $\Sigma^\alpha$. Implicitly, this is what we have done in the argument above.

The same formal structure can also be seen in abstract approaches to BRST-BV theory. As in \cite{BarnichGrigoriev2005a} we can write an odd vector f\/ield $K$
\begin{gather*}
K=\Omega^b_{\;a}\psi^a{\partial\over\partial\psi^b}+U^c_{\;ab}\psi^a\psi^b
{\partial\over\partial\psi^c}+U^d_{\;abc}\psi^a\psi^b\psi^c{\partial\over\partial\psi^d}+\cdots,
\end{gather*}
where $K$ could be either a BRST-operator $Q$ or a BRST-BV operator $s$ depending on the context. In both case, the equation $K^2=0$ generates an sh-Lie algebra.

Contemplating all this one might wonder whether there are more abstract common algebraic structures underlying both the Vasiliev approach and for instance the constructions outlined in Section~\ref{sec:HSBG}, and in particular the BRST-BV approach of \cite{AKHB2007a} (see Section~\ref{sec:RotBRST-BVMA} in the present paper) which superf\/icially look very dif\/ferent. It seems that the language of category theory, and in particular the language of operads, could furnish us with such an abstract framework. Some preliminary remarks on this can be found in Section~\ref{sec:US}. For more connections between the Vasiliev the BRST formalism, see \cite{BarnichGrigorievSemikhatovTipunin2004}, where it is shown that both the free Vasiliev theory and the BRST theory can be derived from a common ``parent f\/ield theory''.

\subsection{Deforming}
The deformation approach starts out with a free gauge theory. We have the action and the Abelian gauge transformations and the object is to deform the action and the gauge transformations by non linear terms, still retaining gauge invariance. This works nicely for spin 1 as was shown a long time ago in \cite{OgievetskyPolubarinov1963}. The analogous deformation approach to spin 2 is considerable more complicated, both technically and conceptually, and there is a long list of classical papers treating this problem \cite{Gupta1952,Gupta1954,Kraichnan1955,Wyss1965,Thirring1961,Feynman1962a,Deser1970,BoulwareDeserKay1979,FangFronsdal1979,Deser1987}. No wonder then, perhaps, that the higher spin interaction problem is even more convoluted. The list of references pertaining to this problem is quite long, for a reasonable subset, see \cite{AKHB2005a}.

\section{Review of the BRST-BV Minkowski approach}\label{sec:RotBRST-BVMA}
As a modern example of the deformation method, let us describe the concrete Fock complex vertex implementation of the BRST-BV approach to higher spin gauge self interactions in MiS in a top-down fashion. The exposition here will be brief, a thorough description can be found in the recent paper \cite{AKHB2007a}, which also contains references to the relevant
theoretical background. For a general discussion of the BV-deformation
approach to interactions see
\cite{BarnichHenneaux1993a,Henneaux1997a}.

Thus we start by writing out the master action to all orders in formal perturbation theory
\begin{gather}\label{eq:MasterAction}
S=\langle\Psi|Q|\Psi\rangle|_{{\rm gh_m}=0}+\sum_{n=3}^\infty g^{n-2}\langle\Psi|^{\otimes n}|{\cal V}_n\rangle|_{{\rm gh_m}=0}.
\end{gather}
Here $|\Psi\rangle$ denotes a formal sum of ghost, f\/ield, antif\/ield, antighost components
\begin{gather*}
|\Psi\rangle=|{\cal C}\rangle\oplus|\Phi\rangle\oplus|\Phi^\#\rangle\oplus|{\cal C}^\#\rangle.
\end{gather*}

The components $|{\cal C}\rangle$, $|\Phi\rangle$, $|\Phi^\#\rangle$, $|{\cal C}^\#\rangle$ live in the Fock complex of mechanics oscillators and ghosts described in Section~\ref{subsec:OscBas}. Actually, the f\/ield $|\Phi\rangle$ is precisely the higher spin f\/ield whereas the f\/ield theory ghost f\/ield $|{\cal C}\rangle$ replaces the gauge parameter $|\Xi\rangle$. The $\#$-decorated components are the corresponding anti-f\/ield and anti-ghost. When we want to refer to the f\/ields without having this concrete realization in mind, we just write $\Psi$.

In this formulation there are no trace constraint operators $T$. Instead, the theory is subject to two global symmetries
\begin{gather*}
\delta_{\cal P}|\Psi\rangle=i\epsilon{\cal P}|\Psi\rangle,\qquad 
\delta_{\cal T}|\Psi\rangle=i\epsilon{\cal T}|\Psi\rangle,
\end{gather*}
where the def\/inition of the operators ${\cal P}$ and ${\cal T}$ can be found in \cite{AKHB2007a}.

It then follows  that $\delta_{\cal P}\langle\Psi|Q|\Psi\rangle=\delta_{\cal T}\langle\Psi|Q|\Psi\rangle=0$. In this formalism there are, for a given primary spin $s$ f\/ield, a further spin $s-2$ f\/ield (corresponding to the trace of the spin $s$ f\/ield) that is not solved for by the trace constraint. For example, the primary spin 5 f\/ield will be accompanied by a new secondary spin 3 f\/ield. It is interesting to note that cubic spin 3 interactions with non-minimal number of derivatives have been found \cite{BekaertBoulangerCnockaert2006a}, i.e.\ with f\/ive derivatives instead of three. This is expected for a secondary spin 3 f\/ield accompanying the primary spin 5 f\/ield.

The vertex operators $|{\cal V}_n\rangle$, encode all interaction data, i.e.\ all $n$-point vertex data as well as all higher order structure function data.

The BRST invariance of the classical theory is now expressed by the master equation \mbox{$(S,S){=}0$} where $(\cdot,\cdot)$ is the anti-bracket of BV theory.

Explicitly calculating $(S,S)=0$ order by order in vertex order $n$ yields
\begin{gather}\label{eq:QV3}
\sum_{r=1}^3 Q_r|{\cal V}_{3}\rangle=0,
\\
\label{eq:QVn}
\sum_{r=1}^{n+1}Q_r|{\cal V}_{n+1}\rangle=-\sum_{p=0}^{\lfloor(n-3)/2\rfloor}|{\cal V}_{p+3}\rangle\diamond|{\cal V}_{n-p}\rangle,
\end{gather}
where $\diamond$ denotes a certain symmetrized contraction in the mechanics Fock complex. These equations encode the structure of a $L_\infty$ algebra as will be shown in Section~\ref{sec:US}.

Extending the global symmetries to the interacting theory yield the following conditions on the vertices
\begin{gather*}
\sum_{r=0}^n{\cal P}_r|{\cal V}_n\rangle=0,\qquad 
\sum_{r=0}^n{\cal T}_r|{\cal V}_n\rangle=0.
\end{gather*}
In order to take into account f\/ield redef\/initions, we introduce f\/ield redef\/inition vertices $|{\cal R}_n\rangle$ and write
\begin{gather*}
|\Psi\rangle\rightarrow|\Psi_r\rangle =|\Psi\rangle +\sum_{n=2}^\infty g^{(n-1)}\langle\Psi|^{\otimes n}|{\cal R}_{n+1}\rangle.
\end{gather*}
Performing f\/ield transformations of this form on the free action produce fake interaction terms of the form
\begin{gather*}
\sum_{n=3}^\infty g^{(n-2)}\langle\Psi|^{\otimes n}\sum_{r=1}^nQ_r|{\cal R}_n\rangle.
\end{gather*}
Comparing to the general form of the interactions given in equation (\ref{eq:MasterAction}), we see that fake interactions can be characterized by
\begin{gather*}
|{\cal V}_n\rangle_{\rm fake}=\sum_{r=1}^nQ_r|{\cal R}_n\rangle,
\end{gather*}
which in the language of homology is to say that fake interaction vertices are exact.

Let us clarify one potentially confusing issue. The vertices $|{\cal V}_n\rangle$ provide us with products of f\/ields, i.e.\ the product of $n$ f\/ields $\Psi_1,\ldots,\Psi_n$, abstractly denoted by ${\bf pr}(\Psi_1,\ldots,\Psi_n)$, can be calculated as
\begin{gather*}
{\bf pr}(\Psi_1,\ldots,\Psi_n)\hookrightarrow\langle\Psi_1|\cdots\langle\Psi_n|{\cal V}_{n+1}\rangle\rightarrow|\Psi_{n+1}\rangle,
\end{gather*}
i.e.\ contracting the $(n+1)$ vertex with $n$ bra f\/ields produces a new ket f\/ield. The index numbers are precisely the numbering of the mechanics Fock spaces including the f\/ield momenta. The vertices enforce momentum conservation. In this context it is natural to chose the abstract products ${\bf pr}$ to be fully symmetric in the abstract f\/ields $\Psi$, and indeed, the Fock f\/ields $|\Psi\rangle$ can be chosen to be Grassmann even.

Now, these products corresponds to what in the mathematics literature are generally denoted by $n$-ary brackets $[\cdot,\cdot,\ldots,\cdot]$, generalizing the two-bracket $[\cdot,\cdot]$ of a dif\/ferential graded Lie algebra. Such a bracket is skew symmetric, rather than symmetric like our 2-products ${\bf pr}(\cdot,\cdot)$ implemented by the three-vertex $|{\cal V}_3\rangle$. This issue is resolved by noting that our product is in fact alternatingly Grassmann odd/even in the mechanics Fock complex due to the presence of $n$ odd vacua $|-\rangle$. Thus the two-product is intrinsically odd, and instead of the conventional grading
\begin{gather*}
{\bf pr}(\Psi_1,\Psi_2) = -(-)^{\epsilon(\Psi_1)\cdot\epsilon(\Psi_2)}{\bf pr}(\Psi_2,\Psi_1),\qquad
\epsilon([\Psi_1,\Psi_2]) = \epsilon(\Psi_1)+\epsilon(\Psi_2),
\end{gather*}
we have
\begin{gather*}
{\bf pr}(\Psi_1,\Psi_2) = (-)^{\epsilon(\Psi_1)\cdot\epsilon(\Psi_2)}{\bf pr}(\Psi_2,\Psi_1),\qquad
\epsilon([\Psi_1,\Psi_2]) = \epsilon(\Psi_1)+\epsilon(\Psi_2)+1,
\end{gather*}
and correspondingly for higher order products. This is also consistent with the recursive equations~(\ref{eq:QVn}) since $\epsilon(Q)=1$.

\section{Unifying structures}\label{sec:US}
It seems quite clear that the algebraic structure of strongly homotopy Lie algebras naturally crop up in various formulations of higher spin gauge theory. How can this be understood from a~formalism independent way? According to the literature on the subject \cite{Stasheff1997a}, $L_\infty$ structures was f\/irst spotted by Stashef\/f in the BBvD analysis~\cite{BerendsBurgersvanDam1985} of the general higher spin interaction problem. Subsequently it was proved that given that the BBvD higher spin gauge algebra exists, it must be sh-Lie~\cite{FulpLadaStasheff2002}. Inf\/luenced by this, and inspired by computer science thinking, I~proved that the sh-Lie structure is grounded already in the syntax of any formal power series formulation of gauge theory. Could this be a hint that sh-Lie algebras, rather than being deep features of higher spins, are just superf\/icial aspects of the formalism? As will be argued in the next section, category theory can throw light on this.

In category theory we do not seek structure by peeking into the objects, but instead build structure on the outside, so to speak. Which is of course precisely what we do when we build models of physical systems. The complex objects so formed can then be re-analyzed, i.e.~peeked into.

\subsection{Abstraction, categories and operads}
In my opinion, and emphasized in \cite{AKHB2005a}, one of the main problems in higher spin theory is to control the inherent complexity. In that paper, we took a rather simpleminded syntax-semantics approach to the problem and were then able to see that the somewhat elusive connections between gauge invariance, BRST-BV formulations and strongly homotopy algebras that has often been referred to in the literature, is grounded already in the syntax of the theory. Now the theory of categories in general and the theory of operads in particular of\/fer a solid mathematical framework for precisely this kind of situation. This should perhaps not come as a surprise since category theory is routinely applied to the syntax-semantics duality in theoretical computer science.

The f\/irst problem one is confronted with upon trying to apply category theory to f\/ield theory~\cite{Baez2004a} is: what are the objects and what are the morphisms? As brief\/ly discussed in \cite{AKHB2007a}, there is no one unique answer to that question, so let us start in another place. Any formulation of interacting higher spin gauge f\/ield theory will involve multi-f\/ield interaction vertices of some sort. The natural categorical correspondence then ought to be the $n$-ary multi-operations ${\cal P}(n)$ of an operad \cite{May1997}. Now the operad in itself only provides the axioms for the ${\cal P}(n)$'s, that is, they provide a syntax for the interactions. To get a concrete model we need a semantics, i.e.\ in computer science terminology: an evaluation, or in category theory language: an algebra for the operad. Indeed an algebra for an operad precisely furnishes an evaluation of the operad. Let us make this more exact.

\subsection{Operads and algebras for operads}
The concept of an operad captures the idea of abstract $n$-ary operations ${\cal P}(n)$ with $n$ input lines and one output line. The ${\cal P}(n)$'s (one set for each $n\geq 1$) can be taken to be vector spaces which is just to say that they can be added and multiplied with numbers from some ground f\/ield $\bf k$. It is then natural to string these $n$-operations together to from new $n$-operations. There are then some issues to contemplate such as associativity of the compositions and permutations of input lines. This is controlled by the axioms of operads (see for example \cite{Leinster}).
	
A potentially confusing issue is the type of objects f\/iguring at the inputs of the $n$-operations. Having f\/ield theory application in mind, there are two basic options. On the one hand, aiming at modeling conf\/iguration space f\/ields, we have just one single object $\Phi$. This then provide input to the $n$-operations and we have the rudiments of a non-polynomial f\/ield theory. On the other hand, aiming at modeling momentum space f\/ields, the objects come labeled by an index $i$ (or $p_i$ in concrete f\/ield theory). Then we have a countable set $\{\Phi_i\}_{i=1}^\infty$ of objects. In this case we have a slight generalization of operads into multi-categories where the $n$-operations take inputs from the set of objects $\{\Phi_i\}$. Specializing to unary operations ${\cal P}(1)$ we have an ordinary category with a countable set of objects and with the ${\cal P}$'s playing the role of arrows.

Another confusing issue is how an abstraction like this can really capture the details of f\/ield theory? How are the n-operations related to the concrete higher spin vertices of the BRST-BV approach? Consider the vector space ${\cal P}(n)$ of $n$-operations. Using the axioms of the multi-category (or the operad) these can be deconstructed into terms of simpler constituents, eventually coming down to a set of basic operations (classically, at the tree level). So, in the set of ${\cal P}(n)$ there is one special element (one for each $n$) which is the abstraction of the concrete vertex operator $|{\cal V}_n\rangle$. Let us denote the corresponding abstract element by ${\cal V}(n)$.

Thus, the objects of the multi-category are abstract f\/ields $\Phi_i$. We then consider the vertex operators $|{\cal V}_n\rangle$ as maps $V_n$ providing evaluations
\begin{gather}\label{eq:Evaluation}
V_n:{\cal V}(n)\otimes \Phi^{\otimes n}\rightarrow\Phi.
\end{gather}
On general non-elementary operations ${\cal P}(n)$, the action of the maps $V_n$ are def\/ined recursively in the standard way of def\/ining maps on recursive data structures.

\subsection{Semantic mapping of the sh-Lie structure}

To see the strength of this simple formalism let us see how the main identities of sh-Lie structure directly maps to the concrete vertex equations (\ref{eq:QV3}), (\ref{eq:QVn}) of Section~\ref{sec:RotBRST-BVMA}.

Let us f\/irst write down the product identities of the sh-Lie algebra as they were derived in~\cite{AKHB2005a} in the case of even abstract f\/ields $\Phi$. Taking into account that the product maps are fully symmetric in all arguments, we can write the product identities as
\begin{gather*}
\sum_{{k=0,l=0}\atop \rm{cycl.\, perm.}}^{k+l=n}{\bf pr}(\Phi^k,{\bf pr}(\Phi^l))=0,
\end{gather*}
where we write the multi-categorical morphism as ${\bf pr}(\Phi^n)$ which we think of as the abstract representation of the product of $n$ f\/ields mapping to a new f\/ield
\begin{gather}\label{eq:ProductMap}
\Phi^{\otimes n}\rightarrow\Phi:{\bf pr}(\Phi_1,\ldots,\Phi_n)\rightarrow\Phi_{n+1}.
\end{gather}
These identities are just special cases of the sh-Lie identities (\ref{eq:SHDefiningEquations}), now written in terms of products instead of brackets.

It will be convenient to introduce a special provision to deal with the one-product ${\bf pr}(\Phi)$ which is naturally interpreted as a linear transformation $K\Phi$. Straining the formalism a little, the one-product can be made to conform to (\ref{eq:ProductMap}), if we write
\begin{gather*}
K_2\Phi_2=K_2\Phi_1\delta_{12}={\bf pr}(\Phi_1)\rightarrow\Phi_2
\end{gather*}
by which we mean that in whatever way we compute the one-product (or linear transformation), we change index on route. This is practical when we do the same in the Fock complex
\begin{gather*}
Q_2|\Phi_2\rangle=Q_1|\Phi_2\rangle\delta_{12}=\langle\Phi_1|{\cal V}_2\rangle.
\end{gather*}
In this way we can think of the action of the BRST operator $Q$ in terms of a 2-vertex $|{\cal V}_2\rangle$.

The semantic map can now be def\/ined. The abstract f\/ields $\Phi$ are simply mapped to Fock complex f\/ibers $|\Phi\rangle$, and the products ${\bf pr}_n$ are mapped to vertices $|{\cal V}_{n+1}\rangle$. Indeed using an arrow~$\hookrightarrow$ to denote the semantic map, we have
\begin{gather*}
\Phi_k\hookrightarrow|\Phi_k\rangle \qquad \mbox{for}\quad  k\in N,\\
{\bf pr}(\Phi_1,\ldots,\Phi_n)\hookrightarrow\langle\Phi_1|\cdots\langle\Phi_n|{\cal V}_{n+1}\rangle \qquad \mbox{for}\quad  n\geq 1 .
\end{gather*}
This is our concrete realization of the {\it evaluation} (\ref{eq:Evaluation}) (in computer science vernacular), or {\it algebra of the operad} as would be the operadic notion.

Note that the last equation precisely utilizes our special provisions for the one-product, i.e.\ the case $n=1$.We are bit lenient with the formalism here, as the $(n+1)$-th Fock space in the vertex $|{\cal V}_{n+1}\rangle$ should really be switched to a bra. As it stands, the last equation produces a ket. It could be f\/ixed at the cost of a more cumbersome notation.

With the ground so prepared we can f\/inally apply the semantic map $\hookrightarrow$ to the left hand side of the product identities to get
\begin{gather*}
\sum_{{k=0,l=0}\atop \rm{cycl.\, perm.}}^{k+l=n}{\bf pr}(\Phi^k,{\bf pr}(\Phi^l)) \hookrightarrow
\sum_{{k=0,l=0}\atop \rm{cycl.\, perm.}}^{k+l=n}\langle\Phi|^{\otimes k}(\langle\Phi|^{\otimes l}|{\cal V}_{l+1}\rangle)\cdot|{\cal V}_{k+2}\rangle.
\end{gather*}
In writing this equation, we are freely switching bra $\leftrightarrow$ ket Fock spaces as need arise to do the contractions. With $l=n-k$ we now have
\begin{gather}\label{eq:ProductIdentityMapped2}
\sum_{{k=0}\atop \rm{cycl. perm.}}^{n-1}\langle\Phi|^{\otimes k}\langle\Phi|^{\otimes n-k}|{\cal V}_{k+2}\rangle\cdot|{\cal V}_{n-k+1}\rangle=0.
\end{gather}
The sum stops at $k=n-1$ since the last term $k=n$ is zero, in the abstract product identities corresponding to ${\bf pr}(\,)=0$, which in the implementation would be $|{\cal V}_1\rangle=0$, i.e.\ there is no 1-vertex.

Then focusing on the f\/irst ($k=0$) and next to last ($k=n-1$) terms, we see, using the conventions introduced for the 2-vertex, that they simply give us
\begin{gather*}
\sum_{r=1}^{n+1}Q_r|{\cal V}_{n+1}\rangle.
\end{gather*}
Then the rest of the terms in (\ref{eq:ProductIdentityMapped2}) pair of\/f nicely in a similar way. Thus the ${\bf pr}(\Phi^k,({\bf pr}\Phi^{n-k}))$ term for $k\geq 1$ maps to precisely ${n\choose k}$ terms containing the vertex combination $|{\cal V}_{k+2}\rangle\cdot|{\cal V}_{n-k+1}\rangle$, while the ${\bf pr}(\Phi^{n-k-1},({\bf pr}\Phi^{k+1}))$ term maps to precisely ${n\choose n-k-1}$ terms also containing the vertex combination $|{\cal V}_{k+2}\rangle\cdot|{\cal V}_{n-k+1}\rangle$. Since ${n\choose k}+{n\choose n-k-1}={n+1\choose n-k}$ we see that we get precisely the correct number of terms to fully symmetrize $|{\cal V}_{k+2}\rangle\cdot|{\cal V}_{n-k+1}\rangle$. This is so because this contraction of vertices has $n+1$ free non-contracted indices (two of the $n+3$ being contracted). Doing the algebra carefully yields
\begin{gather*}
\sum_{k=1}^{\lfloor(n-1)/2\rfloor}|{\cal V}_{k+2}\rangle\diamond|{\cal V}_{n-k+1}\rangle,
\end{gather*}
or, upon re-indexing the sum with $p=k-1$
\begin{gather*}
\sum_{p=0}^{\lfloor(n-3)/2\rfloor}|{\cal V}_{p+3}\rangle\diamond|{\cal V}_{n-p}\rangle.
\end{gather*}
Hence, collecting all the terms, we get
\begin{gather*}
\sum_{r=1}^{n+1}Q_r|{\cal V}_{n+1}\rangle=-\sum_{p=0}^{\lfloor(n-3)/2\rfloor}|{\cal V}_{p+3}\rangle\diamond|{\cal V}_{n-p}\rangle,
\end{gather*}
which is exactly what we got before from the explicit $(S,S)=0$ calculation to all orders in $g$ and antighost number.

In conclusion, the syntactically derived product identities of the sh-Lie algebra maps semantically to equations for the vertices in the Fock complex implementation. This result lends considerable strength to our framework.

If higher spin gauge f\/ields have anything to do with physical reality, then we would eventually have to do numerical calculations. Given the complexity of the theory discerned so far, these calculations would undoubtedly have to be computerized. In that case, all ``objects'' of the theory will have to be mapped to countable inf\/inite data structures (truncated to f\/inite data structures in practice). One might speculate whether in that case formulations  that more easily translate into recursive and algebraic data structures of, for example, a functional programming language, might be more useful than the ordinary ``pen-and-paper'' mathematical formalism of f\/ield theory?

\section{A note on references}
The literature on higher spin f\/ield theory is enormous and growing rapidly. The present paper is not intended to be a full review and leaves out many interesting developments or just mentions them in passing. The reason for this is at least twofold: (i) the limits of my own knowledge, (ii) my wish to put forth a certain point of view, perhaps stressing what has gone un-noticed in other works. Thus I hope the present paper can serve as a complement to other excellent reviews of higher spin gauge theory. My referencing necessarily ref\/lects these limitations and choices. I do apologize for any inadvertent omissions.

\subsection*{Acknowledgements}
Work partly supported by the Research and Education Board at the University College of Bor{\aa}s.

\pdfbookmark[1]{References}{ref}
\LastPageEnding


\begin{thebibliography}{99}

\footnotesize\itemsep=0pt

\bibitem{FangFronsdal1979}
Fang J., Fronsdal C.,
 Deformations of gauge groups: gravitation,
 {\em J. Math. Phys.} {\bf 20} (1979), 2264--2271.

\bibitem{Fronsdal1979conf}
Fronsdal C.,
 Some open problems with higher spins,
 in
  Supergravity, Editors P.~van Nieuwenhuizen and D.Z.~Freedman, North-Holland Publishing Company, 1979, 245--249.

\bibitem{AKHB1985}
 Bengtsson A.K.H.,
 Gauge invariance for spin-3 f\/ields,
 {\em Phys. Rev.~D} {\bf 32} (1985), 2031--2036.

\bibitem{Burgers1985thesis}
 Burgers G.J.H.,
On the construction interactions of f\/ield theories for higher
  spin massless particles,
 PhD Thesis, Rijksuniversteit, Leiden, 1985.

\bibitem{BerendsBurgersvanDam1984}
 Berends F.A., Burgers G.J.H., van Dam H.,
 On spin three self interactions,
 {\em Z. Phys.~C} {\bf 24} (1984), 247--254.

\bibitem{BBB1983a}
 Bengtsson A.K.H., Bengtsson I., Brink L.,
 Cubic interaction terms for arbitrary spin,
 {\em Nuclear Phys.~B} {\bf 227} (1983), 31--40.

\bibitem{FuchsScweigertBook}
Fuchs J., Scweigert C.,
Symmetries, Lie algebras and representations,
 {\it Cambridge Monographs on Mathematical Physics}, Cambridge University
  Press, 1997.

\bibitem{BB1986}
Bengtsson A.K.H., Bengtsson I.,
 Massless higher-spin f\/ields revisited,
 {\em Classical Quantum Gravity} {\bf 3} (1986), 927--936.

\bibitem{Bekaert2007}
Bekaert X.,
 Higher spin algebras as higher symmetries,
\href{http://arxiv.org/abs/0704.0898}{arXiv:0704.0898}.

\bibitem{SiegelZwiebach1987}
Siegel W., Zwiebach B.,
 Gauge string f\/ields from the light-cone,
 {\em Nuclear Phys.~B} {\bf 282} (1987), 125--141.

\bibitem{OuvryStern1987a}
Ouvry S., Stern J.,
 Gauge f\/ields of any spin and symmetry,
 {\em Phys. Lett.~B} {\bf 177} (1987), 335--340.

\bibitem{AKHB1987a}
 Bengtsson A.K.H.,
 A unif\/ied action for higher spin gauge bosons from covariant string
  theory,
 {\em Phys. Lett.~B} {\bf 182} (1987), 321--325.

\bibitem{Meurice1988}
Meurice Y.,
 From points to gauge f\/ields,
 {\em Phys. Lett.~B} {\bf 186} (1988), 189--194.

\bibitem{Labastida1989}
 Labastida J.M.F.,
 Massless particles in arbitrary representations of the {L}orentz
  group,
 {\em Nuclear Phys. B} {\bf 322} (1989), 185--209.

\bibitem{PashnevTsulaia1997}
Pashnev A., Tsulaia M.M.,
 Dimensional reduction and BRST approach to the description of a
  Regge trajectory,
 {\em Modern Phys. Lett.~A} {\bf 12} (1997), 861--870,
 \href{http://arxiv.org/abs/hep-th/9703010}{hep-th/9703010}.

\bibitem{PashnevTsulaia1998a}
Pashnev A., Tsulaia M.,
 Description of the higher massless irreducible integer spins in the
  {BRST} approach,
 {\em Modern Phys. Lett.~A} {\bf 13} (1998), 1853--1864, \href{http://arxiv.org/abs/hep-th/9803207}{hep-th/9803207}.

\bibitem{FranciaSagnotti2002a}
Francia D., Sagnotti A.,
 Free geometric equations for higher spins,
 {\em Phys. Lett.~B} {\bf 543} (2002), 303--310,
 \href{http://arxiv.org/abs/hep-th/0207002}{hep-th/0207002}.

\bibitem{Sundborg2001a}
Sundborg B.,
 Stringy gravity, interacting tensionless strings and massless higher
  spins,
 {\em Nuclear Phys. Proc. Suppl.} {\bf 102} (2001), 113--119,
 \href{http://arxiv.org/abs/hep-th/0103247}{hep-th/0103247}.

\bibitem{Bonelli2003a}
Bonelli G.,
 On the tensionless limit of bosonic strings, inf\/inite symmetries and
  higher spins,
 {\em Nuclear Phys.~B} {\bf 669} (2003), 159--172, \href{http://arxiv.org/abs/hep-th/0305155}{hep-th/0305155}.

\bibitem{SagnottiTsulaia2004a}
Sagnotti A., Tsulaia M.,
 On higher spins and the tensionless limit of string theory,
 {\em Nuclear Phys. B} {\bf 682} (2004), 83--116.
 \href{http://arxiv.org/abs/hep-th/0311257}{hep-th/0311257}.

\bibitem{FranciaSagnotti2006rw}
Francia D., Sagnotti A.,
 Higher-spin geometry and string theory,
 {\em J. Phys. Conf. Ser.} {\bf 33} (2006), 57--72,
 \href{http://arxiv.org/abs/hep-th/0601199}{hep-th/0601199}.

\bibitem{Bianchi2004a}
Bianchi M.,
 Higher spins and stringy ${\rm AdS}_5 \times S_5$,
 {\em Fortsch. Phys.} {\bf 53} (2005), 665--691,
 \href{http://arxiv.org/abs/hep-th/0409304}{hep-th/0409304}.

\bibitem{SagnottiSezginSundell2005}
Sezgin E., Sagnotti A., Sundell P.,
 On higher spins with a strong $sp(2,r)$ condition,
in Proceedings of First Solvay Workshop on Higher-Spin Gauge Theories (May 12--14, 2004, Brussels),
Editors G.~Bonelli, R.~Argurio, G.~Barnich and M.~Grigoriev, Universit\'e Libre de Bruxelles,
International Solvay Institutes for Physics and
Chemistry,
2004, 100--121,
 \href{http://arxiv.org/abs/hep-th/0501156}{hep-th/0501156}.

\bibitem{FranciaMouradSagnotti2007a}
Mourad J., Francia D., Sagnotti A.,
 Current exchanges and unconstrained higher spins,
 {\em Nuclear Phys. B} {\bf 773} (2007), 203--237,
 \href{http://arxiv.org/abs/hep-th/0701163}{hep-th/0701163}.

\bibitem{Maldacena1997a}
Maldacena J.,
 The large $N$ limit of superconformal f\/ield theories and
  supergravity,
 {\em Adv. Theor. Math. Phys.} {\bf 2} (1998), 231--252,
 \href{http://arxiv.org/abs/hep-th/9711200}{hep-th/9711200}.

\bibitem{SezginSundell2002a}
Sezgin E., Sundell P.,
 Massless higher spins and holography,
 {\em Nuclear Phys. B} {\bf 644} (2002), 303--370,
 Erratum, {\em Nuclear Phys. B} {\bf 660} (2003), 403, \href{http://arxiv.org/abs/hep-th/0205131}{hep-th/0205131}.

\bibitem{BeisertBianchiMoralesSamtleben2004a}
Morales J.F., Beisert N., Bianchi M., Samtleben H.,
 Higher spin symmetry and $N = 4$ {SYM},
 {\em J. High Energy Phys.} {\bf 2004} (2004), no.~07, 058, 35~pages, \href{http://arxiv.org/abs/hep-th/0405057}{hep-th/0405057}.

\bibitem{Fronsdal1978}
Fronsdal C.,
 Massless f\/ields with integer spin,
 {\em Phys. Rev.~D} {\bf 18} (1978), 3624--3629.

\bibitem{AKHB2007a}
 Bengtsson A.K.H.,
 Structure of higher spin gauge interactions,
 {\em J. Math. Phys.} {\bf 48} (2007), 072302, 35~pages,
 \href{http://arxiv.org/abs/hep-th/0611067}{hep-th/0611067}.

\bibitem{AKHB1988}
 Bengtsson A.K.H.,
 {BRST} approach to interacting higher-spin gauge f\/ields,
 {\em Classical Quantum Gravity} {\bf 5} (1988), 437--451.

\bibitem{FranciaSagnotti2003a}
Francia D., Sagnotti A.,
 On the geometry of higher-spin gauge f\/ields,
 {\em Classical Quantum Gravity} {\bf 20} (2003), 473--486,
 \href{http://arxiv.org/abs/hep-th/0212185}{hep-th/0212185}.
 
  \bibitem{AKHB1987b}
 Bengtsson A.K.H.,
A one-dimensional invariance principle for gauge f\/ields of
integer
spin, {\em Phys. Lett. B} {\bf 189} (1987), 337--340.

\bibitem{HenneauxTeitelboim1989a}
Henneaux M., Teitelboim C.,
First and second quantized point particles of any spin,
in Quantum
Mechanics of
Fundamental Systems~2, Editors C.~Teitelboim and J.~Zanelli, {\it Series of the Centro de Estudios
Cient{\'i}ficos de
Santiago}, Plenum Press, New York, 1989, 113--152.

\bibitem{EighthNobelSymposium}
Svartholm N. (Editor),
Elementary Particle theory. Relativistic groups and
  analyticity, Almqvist-Wiksell, Wiley Interscience Division, 1968.

\bibitem{CasalbuoniLonghi1975}
Casalbuoni R., Longhi G.,
 A geometrical model for nonhadrons and its implications for hadrons,
 {\em Nouvo Cimento~A} {\bf 25} (1975), 482--502.

\bibitem{Dirac1963}
 Dirac P.A.M.,
 A remarkable represenation of the 3 + 2 de{S}itter group,
 {\em J. Math. Phys.} {\bf 4} (1963), 901--909.

\bibitem{FlatoFronsdal1978a}
Flato M., Fronsdal C.,
 One massless particle equals two {D}irac singletons.
 {\em Lett. Math. Phys.} {\bf 2} (1978), 421--426.

\bibitem{FlatoFronsdal1980a}
Flato M., Fronsdal C.,
 On Dis and Racs,
 {\em Phys. Lett.~B} {\bf 97} (1980), 236--240.

\bibitem{GunaydinSacliogu1982a}
G{\"u}naydin M., Sa{\c c}lio{\= g}lu C.,
 Oscillator-like unitary representations of non-compact groups with a
  {J}ordan structure and the non-compact groups of supergravity,
 {\em Comm. Math. Phys.} {\bf 87} (1982), 159--179.

\bibitem{GunaydinPreprint1989}
G{\"u}naydin M.,
 Singleton and doubleton supermultiplets of space-time supergroups and
  inf\/inite spin superalgebras,  Preprint CERN-TH-5500/89, HU-TFT-89-35,
 1989.


\bibitem{Kibble}
 Kibble T.W.B.,
 Lorentz invariance and the gravitational f\/ield,
 {\em J. Math. Phys.} {\bf 2} (1961), 212--221.

\bibitem{Utiyama}
Utiyama R.,
 Invariant theoretical interpretation of interaction,
 {\em Phys. Rev.} {\bf 101} (1956), 1597--1607.

\bibitem{MacDowellMansouri1977}
 MacDowell S.W., Mansouri F.,
 Unif\/ied geometric theory of gravity and supergravity,
 {\em Phys. Rev. Lett.} {\bf 38} (1977), 739--742.

\bibitem{MansouriChang}
Mansouri F., Chang L.N.,
 Gravitation as a gauge theory,
 {\em Phys. Rev.~D} {\bf 13} (1976), 3192--3200.

\bibitem{Grensing}
Grensing D., Grensing G.,
 General relativity as a gauge theory of the {P}oincar{\'e} group,
 {\em Phys. Rev.~D} {\bf 28} (1983), 286--296.

\bibitem{HehlHeydeKerlick}
 Hehl F.W., von der~Heyde P.,  Kerlick G.D., Nester J.M.,
 General relativity with spin and torsion: foundations and prospects,
 {\em Rev. Modern Phys.} {\bf 48} (1976), 393--416.

\bibitem{KibbleStelle1986}
 Kibble T.W.B., Stelle K.S.,
 Gauge theories of gravity and supergravity,
in Progress in Quantum Field
  Theo\-ry: In Honour of Professor H.~Umewaza, Editors H.~Ezawa and S.~Kamefuchi, North-Holland, Amsterdam, 1986,
  57--81.

\bibitem{StelleWest1980}
 Stelle K.S.,  West P.C.,
 Spontaneously broken de {S}itter symmetry and the gravitational
  holonomy group,
 {\em Phys. Rev. D} {\bf 21} (1980), 1466--1488.

\bibitem{BekaertCnockaertIazeollaVasiliev2005}
Iazeolla C., Bekaert X., Cnockaert S., Vasiliev M.A.,
Nonlinear higher spin theories in various dimensions,
in Proceedings of First Solvay Workshop on Higher-Spin Gauge Theories (May 12--14, 2004, Brussels),
Editors G.~Bonelli, R.~Argurio, G.~Barnich and M.~Grigoriev, Universit\'e Libre de Bruxelles,
International Solvay Institutes for Physics and
Chemistry,
2004, 132--197, \href{http://arxiv.org/abs/hep-th/0503128}{hep-th/0503128}.



\bibitem{JohanEngqvist}
Engqvist J.,
Dualities, symmetries and unbroken phases in string theory,
 PhD Thesis, Uppsala University, 2005.

\bibitem{DAuriaFre1982a}
D'Auria R., Fr{\'e} P.,
 Geometric supergravity in $D=11$ and its hidden supergroup,
 {\em Nuclear Phys.~B} {\bf 201} (1982), 101--140, Errata, {\em Nuclear Phys.~B} {\bf 206} (1982), 496.

\bibitem{Sullivan1977}
Sullivan D.,
 Inf\/initesimal computations in topology,
 {\em Inst. Hautes \'Etudes Sci. Publ. Math.} (1977), no.~47,  269--331.

\bibitem{Vasiliev2005a}
 Vasiliev M.A.,
 Actions, charges and of\/f-shell f\/ields in the unfolded dynamics
  approach,
 {\em Int. J. Geom. Methods Mod. Phys.} {\bf 3} (2006), 37--80,
 \href{http://arxiv.org/abs/hep-th/0504090}{hep-th/0504090}.

\bibitem{LadaStasheff1993a}
Lada T., Stashef\/f J.,
 Introduction to {sh-Lie} algebras for physicists,
 {\em Internat. J. Theoret. Phys.} {\bf 32} (1993), 1087--1104, \href{http://arxiv.org/abs/hep-th/9209099}{hep-th/9209099}.

\bibitem{LadaMarkl1995a}
Lada T., Markl M.,
 Strongly homotopy {L}ie algebras,
 {\em Comm. Algebra} {\bf 23} (1995), 2147--2161, \href{http://arxiv.org/abs/hep-th/9406095}{hep-th/9406095}.

\bibitem{Voronov2003a}
Voronov T.,
 Higher derived brackets and homotopy algebras,
{\it J. Pure Appl. Algebra} {\bf 202} (2005), 133--153,
\href{http://arxiv.org/abs/math.QA/0304038}{math.QA/0304038}.


\bibitem{AzcrragaBueno1997a}
 de~Azc{\'a}rraga J.A., P{\'e}rez Bueno J.C.,
 Higher-order simple {L}ie algbras,
 {\em Comm. Math. Phys.} {\bf 184} (1997), 669--681, \href{http://arxiv.org/abs/hep-th/9605213}{hep-th/9605213}.

\bibitem{BarnichGrigoriev2005a}
Barnich G., Grigoriev M.,
 BRST extension of the non-linear unfolded formalism,
 \href{http://arxiv.org/abs/hep-th/0504119}{hep-th/0504119}.

\bibitem{BarnichGrigorievSemikhatovTipunin2004}
Semikhatov A., Barnich G., Grigoriev M., Tipunin I.,
 Parent f\/ield theory and unfolding in {BRST} f\/irst-quantized terms,
 {\em Comm. Math. Phys.} {\bf 260} (2005), 147--181,
\href{http://arxiv.org/abs/hep-th/0406192}{hep-th/0406192}.

\bibitem{OgievetskyPolubarinov1963}
 Ogievetski V.I., Polubarinov I.V.,
 Interacting f\/ields of spin 1 and symmetry properties,
 {\em Ann. Phys.} {\bf 25} (1963), 358--386.

\bibitem{Gupta1952}
 Gupta S.N.,
 Quantization of {E}instein's gravitational f\/ield: linear
  approximation,
 {\em Proc. Phys. Soc. Sect.~A} {\bf 65} (1952), 161--169.

\bibitem{Gupta1954}
 Gupta S.N.,
 Gravitation and electromagnetism,
 {\em Phys. Rev. (2)} {\bf 96} (1954), 1693--1685.

\bibitem{Kraichnan1955}
 Kraichnan R.H.,
 Special-relativistic derivation of generally covariant gravitation
  theory,
 {\em Phys. Rev.} {\bf 98} (1955), 1118--1122.

\bibitem{Wyss1965}
Wyss W.,
 Zur unizit{\"a}t der {G}ravitationstheorie,
 {\em Helv. Phys. Acta} {\bf 38} (1965), 469--480.

\bibitem{Thirring1961}
Thirring W.E.,
 An alternative approach to the theory of gravitation,
 {\em Ann. Phys.} {\bf 16} (1961), 96--117.

\bibitem{Feynman1962a}
 Feynman R.P.,
Feynman lectures on gravitation,
 Westview Press, 2002. 

\bibitem{Deser1970}
Deser S.,
 Self-interaction and gauge invariance,
 {\em Gen. Relativity Gravitation} {\bf 1} (1970), 9--18.

\bibitem{BoulwareDeserKay1979}
 Boulware D.G., Deser S., Kay J.H.,
 Supergravity from self-interaction,
 {\em Phys.~A} {\bf 96} (1979), 141--162.

\bibitem{Deser1987}
Deser S.,
 Gravity from self-interaction in a curved background,
 {\em Classical Quantum Gravity} {\bf 4} (1987), L99--L105.

\bibitem{AKHB2005a}
 Bengtsson A.K.H.,
 An abstract interface to higher spin gauge f\/ield theory,
 {\em J. Math. Phys.} {\bf 46} (2005), 042312, 23~pages,
 \href{http://arxiv.org/abs/hep-th/0403267}{hep-th/0403267}.
 
 \bibitem{BarnichHenneaux1993a}
Barnich G., Henneaux M.,
Consistent couplings between f\/ields with a gauge freedom and
deformations of the master equation,
{\em Phys. Lett. B} {\bf 311} (1993), 123--129,
\href{http://arxiv.org/abs/hep-th/9304057}{hep-th/9304057}.

\bibitem{Henneaux1997a}
Henneaux M.,
Consistent interactions between gauge f\/ields: the
cohomological
approach,
in   Secondary Calculus and Cohomological Physics,  Editors M.~Henneaux, J.~Krasil'shchik and A.~Vinogradov,
  {\it Contemp. Math.} {\bf 219} (1998), 93--109, \href{http://arxiv.org/abs/hep-th/9712226}{hep-th/9712226}.

\bibitem{BekaertBoulangerCnockaert2006a}
Boulanger N., Bekaert X., Cnockaert S.,
 Spin three gauge theory revisited,
 {\em J. High Energy Phys.} {\bf 2006} (2006), no.~01, 052, 34~pages,
 \href{http://arxiv.org/abs/hep-th/0508048}{hep-th/0508048}.

\bibitem{Stasheff1997a}
Stashef\/f J.,
 The (secret?) homological algebra of the {B}atalin--{V}ilkovisky
  approach,
in   Secondary Calculus and Cohomological Physics,  Editors M.~Henneaux, J.~Krasil'shchik and A.~Vinogradov,
  {\it Contemp. Math.} {\bf 219} (1998), 195--210, \href{http://arxiv.org/abs/hep-th/9712157}{hep-th/9712157}.

\bibitem{BerendsBurgersvanDam1985}
Berends F.A., Burgers G.J.H., van Dam H.,
 On the theoretical problems in constructing interactions invol\-ving
  higher-spin massless particles,
 {\em Nuclear Phys.~B} {\bf 260} (1985), 295--322.



\bibitem{FulpLadaStasheff2002}
Fulp R., Lada T., Stashef\/f J.,
 {Sh-Lie} algebras induced by gauge transformations,
 {\em Comm. Math. Phys.} {\bf 231} (2002), 25--43,
\href{http://arxiv.org/abs/math.QA/0012106}{math.QA/0012106}.

  \bibitem{Baez2004a}
Baez J.,
 Quantum quandaries: a category-theoretic perspective, in
Structural Foundations of Quantum Gravity, Editors
 D.P.~Rickles, S.R.D.~French and J.~Saatsi,   Oxford University Press,
2006, 240--265, \mbox{\href{http://arxiv.org/abs/quant-ph/0404040}{quant-ph/0404040}}.

\bibitem{May1997}
 May J.P.,
 Operads, algebras and modules,
 in Operads: Proceedings of Renaissance Conferences (1995, Hartford, CT/Luminy),
 {\em Contemp. Math.} {\bf 202} (1997), 15--31.

\bibitem{Leinster}
Leinster T.,
Higher operads, higher categories,
 Cambridge University Press, 1989.




\end{thebibliography}
\end{document}